\documentclass{ckm}                 
\usepackage{amssymb}

\usepackage{txfonts}            

\confname{Workshop on the CKM Unitarity Triangle, IPPP Durham, April
  2003}

\title{Application of the heavy quark expansion:  $|V_{ub}|$ and spectral
moments}

\author{M Luke}
\address{Department of Physics, University of Toronto}

\def\lqcd{\Lambda_{\rm QCD}}

\def\qcut{q_0}

\def\OMIT#1{{}}

\def\qcut{q_{\rm cut}}
\def\mxcut{m_{\rm cut}}

\def\gcut{G(\qcut^2,\mxcut)}
\def\gev{{\rm GeV}}
\def\mev{{\rm MeV}}

\def\mbups{m_b^{1S}}
\def\d{{\rm d}}

\def\vereq#1#2{\lower3pt\vbox{\baselineskip1pt\lineskip1pt
    \ialign{\\$#1\hfill##\hfil\\$\crcr#2\crcr\sim\crcr}}}

\begin{document}

\begin{abstract} The use of inclusive $B$ decays to determine $|V_{ub}|$,
$|V_{cb}|$, $m_b$ and heavy quark matrix elements via spectral moments is
discussed.
\end{abstract}

\maketitle


\section{$|V_{ub}|$}

A precise and model independent determination of the magnitudes of the
Cabibbo-Kobayashi-Maskawa
(CKM) matrix elements $V_{ub}$ is important for testing the Standard Model at
$B$ factories via the comparison of the angles and the sides of the unitarity
triangle.  

$|V_{ub}|$ is notoriously hard to measure model-independently.  The first
extraction of $|V_{ub}|$ from experimental data relied on
a study of the lepton energy spectrum in inclusive charmless semileptonic $B$
decay~\cite{CLEO1}, a region in which (as will be discussed) the rate is highly
model-dependent.  $|V_{ub}|$ has also been measured from exclusive
semileptonic $B\to \rho \ell \bar\nu$ and $B\to \pi \ell \bar\nu$
decay~\cite{CLEOexcl}.  These exclusive determinations also suffer from model
dependence, as they rely on form factor models (such as light-cone sum rules
\cite{lcsr}) or quenched lattice calculations at the present time (for a review
of recent lattice results, see \cite{kron}).  

In contrast, inclusive decays are quite simple theoretically, and
if it were not for the huge background from decays to charm, it would be
straightforward to determine $|V_{ub}|$ from inclusive semileptonic decays. 
Inclusive $B$ decay rates can be computed model independently in a series in
$\lqcd/m_b$ and $\alpha_s(m_b)$ using the heavy quark expansion (HQE)
~\cite{CGG,incl,MaWi,Blok}. 
At leading order, the $B$ meson decay rate is equal to the $b$ quark decay rate.
The leading nonperturbative corrections of order $\lqcd^2 / m_b^2$ are
characterized by two heavy quark effective theory (HQET) matrix elements,
usually called $\lambda_1$ and $\lambda_2$.  These matrix elements also occur in
the expansion of the $B$ and $B^*$ masses in powers of $\lqcd/m_b$,
\begin{equation}\label{massrelation}
m_{B(B^*)} = m_b + \bar\Lambda 
  - {\lambda_1 + 3(-1)\lambda_2 \over 2m_b} + \ldots\ .
\end{equation}
Similar formulae hold for the $D$ and $D^*$ masses.  The parameters
$\bar\Lambda$ and $\lambda_1$ are independent of the heavy $b$ quark mass,
while there is a weak logarithmic scale dependence in $\lambda_2$.  The
measured $B^*-B$ mass splitting fixes $\lambda_2(m_b) = 0.12\,{\rm GeV}^2$,
while $m_b$ (or, more precisely, a well-defined
short-distance mass such as $m_b^{(1S)}$ \cite{upsexp,breview}) and $\lambda_1$
may be determined from other physical
quantities (as will be discussed in the second part of this talk).   Since the
parton level decay
rate is proportional to $m_b^5$, the uncertainty in $m_b$ is a dominant source
of uncertainty
in the relation between $B\rightarrow X_u \ell\bar\nu$ and $|V_{ub}|$; an
uncertainty in $m_b$ 
of $50$ MeV corresponds to a $\sim5\%$ determination of
$|V_{ub}|$\cite{upsexp,burels}.

Unfortunately, the $B\to X_u\ell\bar\nu$ rate can only be measured imposing
severe cuts on the phase space to eliminate the $\sim 100$ times larger $B\to
X_c\ell \bar\nu$ background.  Since the predictions of the OPE are only model
independent for sufficiently inclusive observables, these cuts can destroy
the convergence of the expansion.  This is the case for two kinematic regions 
for which the charm background is absent and which have received much attention:
the large lepton energy region,  $E_\ell > (m_B^2-m_D^2)/2m_B$, and the small
hadronic invariant mass region, $m_X < m_D$ \cite{BKP,FLW,DU,delphi00}.

The poor behaviour of the OPE for these quantities is slightly subtle, because
in both cases there is sufficient phase space for many different resonances to
be produced in the final state, so an inclusive description of the decays is
still appropriate. However, in
both of these regions of phase space the $\bar
B\to X_u\ell\bar\nu$ decay products are dominated by high energy, low invariant
mass hadronic states, 
\begin{equation}\label{shapefnregime} 
E_X\sim m_b,\ m_X^2\sim m_D^2\sim\lqcd m_b\ll m_b^2.
\end{equation} 
In this region the differential rate is very sensitive to the details of the
wave function of the $b$ quark in the $B$ meson \cite{shape}.  This can be seen
simply from the kinematics.
A $b$ quark in a $B$ meson
has momentum
\begin{equation}
p_b^\mu=m_b v^\mu+k^\mu
\end{equation}
where $v^\mu$ is the four-velocity of the quark, and $k^\mu$ is a small residual
momentum of order $\lqcd$.  If the momentum transfer to the final state leptons
is $q$, the invariant mass of the final state hadrons is
\begin{eqnarray}\label{kinexp}
m_X^2&=&(m_b v+k-q)^2=(m_b v-q)^2\\
&+&2k\cdot(m_b v-q)+O(\lqcd^2).\nonumber
\end{eqnarray}
Over most of phase space, the second term is suppressed relative to the first by
one power of $\lqcd/m_b$, and so may be treated as a perturbation.  This
corresponds to the usual OPE.  However, in the region (\ref{shapefnregime})
$E_X$ is large and $m_X$ is small, $m_b v^\mu-q^\mu=(E_X,0,0,E_X)+O(\lqcd)$ is
almost light-like, and the first two terms are the same order,
\begin{equation}
m_X^2=(m_b v-q)^2+2 E_X k_++\dots,\ k_+\equiv k_0+k_3 .
\end{equation}
The differential rate in this ``shape function region" is therefore sensitive at
leading order to the wave function $f(k_+)$ which describes the distribution of
the light-cone component of the residual momentum of the $b$ quark.
$f(k_+)$ is a nonperturbative function and cannot be calculated  analytically,
so the rate in the region (\ref{shapefnregime}) is model-dependent even at
leading order in $\lqcd/m_b$.

\OMIT{In terms of the OPE, this light-cone wave function arises because of
subleading
terms in the OPE proportional to
$E_X\lqcd/m_X^2$, which are suppressed over most of phase space
but are $O(1)$ in the region (\ref{shapefnregime}).}
Near the endpoint, the OPE for the $E_\ell$ spectrum has the form (where $y=2
E_\ell/m_b$)
\begin{eqnarray}\label{singularOPE}
{d\Gamma\over dy}&\sim& \nonumber\\
&&\hskip-1cm 2\theta(1-y)-{\lambda_1\over 3
m_b^2}\delta^\prime(1-y)-{\rho_1\over 9
m_b^3}\delta^{\prime\prime}(1-y)+\dots\nonumber\\
&-&{\lambda_1\over 3 m_b^2}\delta(1-y)-{11\lambda_2\over
m_b^2}\delta(1-y)+\dots\ \nonumber\\
&+&\dots
\end{eqnarray}
For $1-y\sim\lqcd/m_b$, the terms on the first line are all parametrically
$O(1)$, and so must be summed to all orders, and the result may be written as a
convolution of $f(k_+)$ with the parton-level rate \cite{shape}.  The terms of
the second line are less singular near $y=1$, and so correspond to subleading
effects, which will be discussed later. 

\begin{figure*}[tbh]
\includegraphics*[width=15cm]{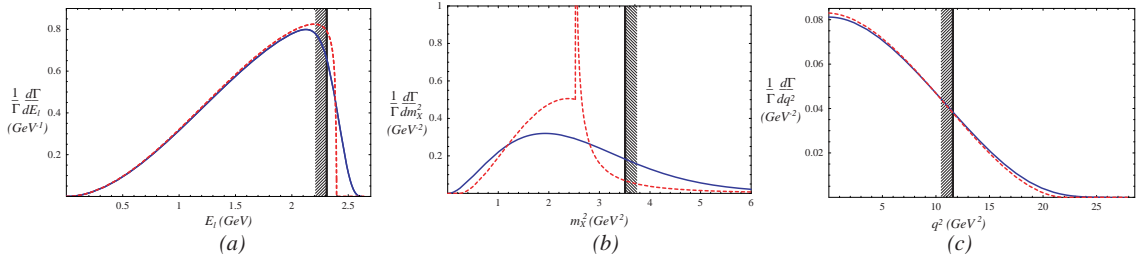}
\caption{The shapes of the lepton energy, hadronic invariant mass and
leptonic invariant mass spectra.
The dashed curves are the $b$ quark decay results to ${\cal O}(\alpha_s)$,
while the solid curves are obtained by convoluting the parton-level rate with
the model distribution
function $f(k_+)$ in Eq.~(\ref{sfn}).  The unshaded side of the vertical lines
indicate the region free from charm background.}
\label{threespectra}
\end{figure*}

The situation is illustrated in Fig.~\ref{threespectra}(a-b), 
where the lepton energy and hadronic invariant mass spectra are plotted in the
parton model (dashed curves) and incorporating a simple one-parameter
model for the distribution function (solid curves)~\cite{dFN}
\begin{equation}\label{sfn}
f(k_+)={32\over \pi^2 \Lambda}\, (1-x)^2\, 
  e^{-{\textstyle{4\over \pi}}(1-x)^2} \theta(1-x)
\end{equation}
where
\begin{equation}
x \equiv {k_+\over \Lambda} \,, \qquad \Lambda=0.48\,{\rm GeV}\,.
\end{equation}
The differences between the curves in the regions of interest indicate the
sensitivity of the spectrum to the precise form of $f(k_+).$
In both curves, the unshaded side of the vertical line denotes the region free
from charm background.  Because $m_D^2\sim \lqcd m_B$, the integrated rate in
this region is very sensitive to the form of $f(k_+)$, complicating the issue of
determining $|V_{ub}|$ model-independently.

\subsection{Optimized Cuts}

One solution to the problem of sensitivity to nonperturbative effects is 
to find a set of cuts which eliminate the charm background
but do not destroy the convergence of the OPE, so that the distribution function
$f(k_+)$ is not required.  In Ref.~\cite{BLL} it was pointed out that this is
the
situation for a cut on the dilepton invariant mass.  Decays with 
\begin{equation}\label{qsqcutregion}
q^2 > (m_B - m_D)^2
\end{equation}
must arise from $b\to u$ transition.  Such a cut
forbids the hadronic final state from moving fast in the $B$ rest frame, and
simultaneously
imposes $m_X < m_D$ and
$E_X < m_D$.  Thus, the light-cone expansion which gives rise to the shape
function is not relevant in this region of phase space~\cite{DU,BI}.   The
effect of convoluting the $q^2$ spectrum with the model distribution function in
Eq.~(\ref{sfn}) is illustrated in Fig.~\ref{threespectra}(c).
The region selected by a $q^2$ cut is entirely contained
within the $m_X^2$ cut, but because the dangerous region
of high energy, low invariant mass final states is not included, the OPE
does not break down.  The price to be paid is that the relative size of the
unknown
$\lqcd^3/m_b^3$ terms in the OPE grows as the $q^2$ cut is raised. 
Equivalently, as was stressed in
\cite{Matthias},  the effective expansion parameter
for integrated rate inside the region (\ref{qsqcutregion}) is $\lqcd/m_c$, not
$\lqcd/m_b$.  In addition, the integrated cut rate is very sensitive to $m_b$,
with a $\pm 80\,\mev$ error in $m_b$ corresponding to a $\sim \pm 10\%$
uncertainty in $|V_{ub}|$ 
\cite{Matthias,doublecut}.

A further important source of uncertainty arises from weak annihilation (WA)
graphs \cite{WA}.  WA arises at $O(\lqcd^3/m_b^3)$
in the OPE, but is enhanced by a factor of $\sim16\pi^2$ because there are only
two particles in the final state compared with $b\to u\ell\bar\nu_\ell$. 
Because WA contributes only at the endpoint of the $q^2$ spectrum, it is
independent of $\qcut^2$ and $\mxcut$:
\OMIT{\begin{eqnarray}\label{WAcont}
{\d\Gamma_{WA}\over \d q^2}&=&-{2 G_F^2 |V_{ub}|^2 m_b^2\over
3\pi}\delta(q^2-m_b^2)\nonumber\\
&&\times {1\over 2 m_B}\langle B|O^u_{V-A}-O^u_{S-P}|B\rangle
\end{eqnarray}}
\begin{equation}\label{WAcont}
{\d\Gamma_{WA}\over \d q^2}\sim (B_2-B_1)\delta(q^2-m_b^2).
\end{equation}
$B_1$ and $B_2$ are matrix elements which are equal for both charged and neutral
$B$'s under the factorization hypothesis, and so the size
of the WA effect depends on the size of factorization violation.  
\OMIT{A simple estimate gives
\begin{equation}
\delta\gcut\sim 0.03\left({f_B\over
0.2\,\gev}
\right)^2\left({B_2-B_1\over 0.1}\right).
\end{equation}}
Assuming factorization is violated at the $10\%$ level gives a corresponding
uncertainty in $|V_{ub}|$ from a pure $q^2$ cut of $\sim 10\%$ \cite{WA}; 
however, this estimate is
highly uncertain, being proportional to $16\pi^2\times\mbox{(factorization\
violation)}$.  In addition, since the contribution is fixed at maximal $q^2$,
the corresponding uncertainty grows as the cuts are tightened, reducing the
integrated rate.

These uncertainties may be reduced by considering more complicated kinematic
cuts: in \cite{doublecut} it was proposed that by combining cuts on both the
leptonic and hadronic invariant masses the theoretical uncertainty on $|V_{ub}|$
could be
minimized.  For a fixed cut on $m_X$, lowering the bound on $q^2$
increases the cut rate and decreases the relative size of the $1/m_b^3$ terms
(including the WA terms), while only introducing a small dependence on $f(k_+)$.
 Since this
dependence is so weak, a crude measurement of $f(k_+)$ suffices to keep the
corresponding theoretical error negligible.  The sensitivity to $m_b$ is also
reduced.

\begin{figure}[t]
\includegraphics*[width=7cm]{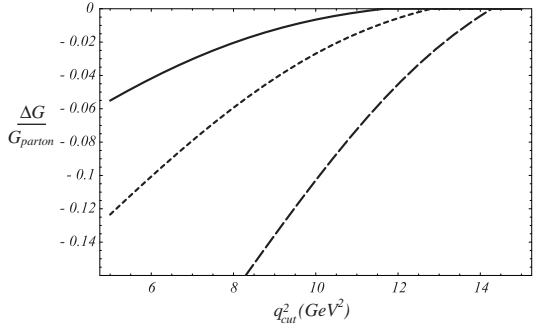}
\caption{The effect of the model structure function (\ref{sfn})
on $\gcut$ as a function of
$\qcut^2$ for $\mxcut = 1.86\,\gev$ (solid line), $1.7\,\gev$ (short
dashed line) and $1.5\,\gev$ (long dashed line).}
\label{Sfig}
\end{figure}

Defining the function $\gcut$ by
\begin{eqnarray}\label{defineg}
&&\Gamma(q^2<q^2_{\rm cut}, m_X<m_{\rm cut})
\equiv \nonumber\\
&&\qquad{G_F^2 |V_{ub}|^2\, (4.7\,\gev)^5\over 192\pi^3}\; \gcut\,,
\end{eqnarray}
the dependence of $\gcut$ on $f(k_+)$ for various cuts is illustrated in 
Fig.\ \ref{Sfig}.   The estimated
uncertainty from other sources is given for a variety of cuts in Table
\ref{finaltable}.  Since the uncertainty in $|V_{ub}|$ is half that of $\gcut$,
we see that, depending on the cuts, theoretical errors at the $5-10\%$ level are
possible.

\begin{table*}[hbt]
\caption{$\gcut$, as defined in Eq.~(\ref{defineg}), for several different
choices 
of $(\qcut^2,\mxcut)$, along with the uncertainties.  The
fraction of $B\to X_u\ell\bar\nu$ events included by the cuts is $1.21\,\gcut$.
$\Delta_{\rm
struct}G$ gives the fractional  effect of the structure function $f(k_+)$ in
the simple model (\ref{sfn}); we do not include an uncertainty on
this in our error estimate. The overall uncertainty $\Delta G$ is obtained
by combining the other uncertainties in quadrature.   The two values correspond
to $\Delta \mbups=\pm 80\,\mev$ and $\pm 30\,\mev$. The uncertainty in
$|V_{ub}|$ is half of $\Delta G$. }

\begin{tabular}{|c||c|c|ccc|c|}\hline
Cuts on $(q^2,\,m_X)$  &  ~$\gcut$~
   &  ~$\Delta_{\rm struct}G$~ &  ~$\Delta_{\rm pert}G$  &  
   $\matrix{\Delta_{m_b}G \cr \pm80/30\,\mev}$  &
   $\Delta_{1/m^3} G$~   &  $\Delta G$  \\ \hline\hline
\multicolumn{1}{|c}{Combined cuts} & \multicolumn{6}{c|}{} \\ \hline
$6\,\gev^2, 1.86\,\gev$ &  0.38  &  $ -4\%$  &4\%  &  13\%/5\%  &  6\%  &
15\%/9\%  \\
$8\,\gev^2, 1.7\,\gev$ &  0.27 &  $-6\%$&  6\%  &  15\%/6\%  &  8\%  &
18\%/12\% \\ 
$ 11\,\gev^2, 1.5\,\gev$  & 0.15 &  $-7\%$ &13\% & 18\%/7\% & 16\% &
27\%/22\% \\ \hline\hline
\multicolumn{1}{|c}{Pure $q^2$ cuts} & \multicolumn{6}{c|}{} \\ \hline
~$(m_B-m_D)^2, m_D$~ & 0.14 & --\,--&15\% &19\%/7\% & 18\% & 
~30\%/24\%~ \\
$(m_B-m_{D^*})^2, m_{D^*}$& 0.17 & --\,--&13\% &17\%/7\% & 14\% &26\%/20\%\\
\hline
\end{tabular}\\[4pt]
\label{finaltable}
\end{table*}

\subsection{$f(k_+)$ and subleading corrections}

Alternatively, one can reduce the theoretical uncertainty in $|V_{ub}|$ by
measuring the universal structure function $f(k_+)$ in some other process
\cite{shape,shape2}.
The best way to measure the structure function $f(k_+)$ is from the photon
energy spectrum of the inclusive decay $B \to X_s \gamma$. Up to perturbative
and subleading twist corrections, this spectrum is directly proportional to the
structure function,
\begin{eqnarray}\label{diffEgamma}
\frac{d\Gamma}{dE_\gamma} &=&  \frac{G_F^2 |V_{tb}V_{ts}^*|^2 \alpha |C_7^{\rm
eff}|^2m_b^5}{32 \pi^4} f(E_\gamma)\,.
\end{eqnarray}
Thus, combining data on $B \to X_s \gamma$ with data from $B \to X_u \ell \bar
\nu$, one can eliminate the dependence on the structure function and therefore
determine $|V_{ub}|$ with no model dependence at leading order \cite{shape,LLR}.
At tree level, the relation is
\begin{equation}\label{Vubrelation}
\left|\frac{V_{ub}}{V_{tb}V_{ts}^*}\right| =
\left( \frac{3 \alpha}{\pi} |C_7^{\rm eff}|^2
\frac{\Gamma_u(E_c)}{\Gamma_s(E_c)}\right)^\frac12
\left(1 + \delta(E_c)\right)
\end{equation}
where 
\begin{eqnarray}\label{GuGsdefs}
\Gamma_u(E_c) &\equiv& \int_{E_c}^{m_B/2} d E_\ell \frac{d \Gamma_u}{d E_\ell}\\
\Gamma_s(E_c) &\equiv& \frac{2}{m_b} \int_{E_c}^{m_B/2} d E_\gamma (E_\gamma -
E_c)\frac{d \Gamma_s}{d E_\gamma} \,\nonumber
\end{eqnarray}
and $\delta(E_c)$ contains terms suppressed by $O(\lqcd/m_b)$.  An analogous
relation holds for the hadronic invariant mass spectrum \cite{FLW,DU}.  In
addition to higher twist effects, there are perturbative corrections to
(\ref{Vubrelation}).  Most important of these are the parametrically large
Sudakov logarithms, which have been summed to subleading order \cite{LLR}.  In
addition, contributions from additional operators which contribute to $B\to
X_s\gamma$ have been calculated  \cite{mn01}.  The CLEO collaboration
\cite{cleovub} recently used a variation of this approach to determine
$|V_{ub}|$ from their measurements of the $B\to X_s\gamma$ photon spectrum and
the charged lepton spectrum in $B\to X_u\ell\bar\nu_\ell$.
\OMIT{
\begin{equation}
|V_{ub}|=(4.08\pm 0.34\pm 0.44\pm 0.16\pm 0.24)\times 10^{-3}
\end{equation}
where the first two errors are experimental, and the latter two estimates of the
theoretical uncertainty.  (As discussed below, the theoretical uncertainty is
probably underestimated here.)}

The subleading corrections to (\ref{Vubrelation}) contained in $\delta(E_c)$
have only recently been studied \cite{blm01,blm02,llw02,mn02}.  These are
analogous to higher twist effects in DIS, and there are two separate effects,
each of which is large.

\begin{figure}[tbh]
\includegraphics[width=7cm]{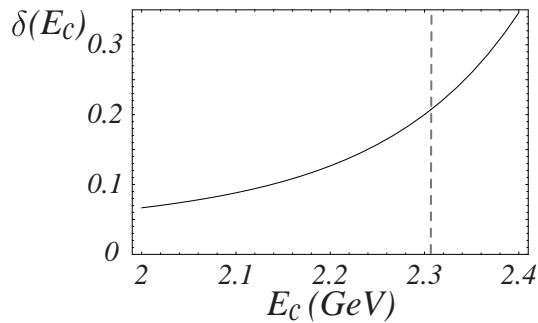}
\caption{The subleading twist corrections to the $|V_{ub}|$ relation
(\ref{Vubrelation}) as a function on the lepton energy cut, using the simple
model described in the text, from Ref.\ \cite{blm02}. The vertical line
corresponds to the kinematic endpoint of the semileptonic $b \to c$
decay.\label{fig2}}
\end{figure}

First of all, the $O(1/m_b)$ corrections happen to have a large numerical
prefactor.  This is easiest to see by looking at the OPE for the lepton spectrum
in semileptonic $b\to u$ decay, Eqn.\ (\ref{singularOPE}).  The terms in the
first line are universal, and so are the same for $B\to X_s\gamma$ and $B\to
X_u\ell\bar\nu_\ell$ decays. The subleading terms on the second line are not
universal, and sum to subleading distribution functions \cite{blm01}.  Note that
the subleading $\lambda_2$ term has a coefficient of 11, whereas the
corresponding coefficient in the $B\to X_s\gamma$ is 3, giving an $O(\lqcd/m_b)$
correction which is enhanced by a factor of 8 over the na\"\i ve dimensional
estimate.  Since this is just the first term of an infinite series, one cannot
immediately determine the size of the subleading correction, but for a simple
model the corresponding shift of $|V_{ub}|$ is plotted in Fig.~\ref{fig2}.  For
a charged lepton cut of $2.3\,\gev$, this corresponds to a $\sim 15\%$ shift in
the extracted value of $|V_{ub}|$.    While this is a substantial shift, it was
argued in \cite{mn02} that the magnitude of this shift is quite insensitive to
the model chosen for the subleading distribution function, and so the
corresponding uncertainty in $|V_{ub}|$ is much smaller than the overall shift
of $\sim 15\%$. 

A second source of uncertainty arises because of the WA graphs discussed in the
previous section.  In the region near $y=1$, the WA graph is the first term of
an infinite series which resums into a sub-subleading (relative order
$(1/m_b^2)$) distribution function \cite{llw02}.  As before, the size of the WA
contribution is difficult to determine reliably; the authors of \cite{llw02}
estimate the corresponding uncertainty in $|V_{ub}|$ to be at the $\sim 10\%$
level (with unknown sign) for a cut $E_\ell>2.3\,\gev$.  For both subleading
effects, the fractional uncertainty in $|V_{ub}|$ is reduced considerably as the
cut on $E_\ell$ is lowered below 2.3\,\gev.  

\subsection{Summary for $|V_{ub}|$}

We are left in the fortunate situation of having a number of theoretically clean
methods of determining $|V_{ub}|$ from inclusive decays, each of which has
advantages and disadvantages which are summarized in Table \ref{comparetable}.  
Because the different techniques have different sources of uncertainty,
agreement between these methods (combined with future unquenched lattice
predictions for $B\to\pi\ell\bar\nu$ decays) will give evidence that the
different sources of uncertainty have been correctly estimated.

In the future, experimental measurements can help reduce the theoretical errors
in a number of ways:
\begin{itemize}
\item better determinations of $m_b$ (through, for example, moments of $B$ decay
distributions) can reduce the largest single source of uncertainty for
determinations using optimized cuts,
\item the size of WA effects may be tested by comparing $D^0$ and $D_S$
semileptonic decays, or by extracting $|V_{ub}|$ from $B^\pm$ and $B^0$ decays
separately,
\item improved measurements of the $B\to X_s\gamma$ spectrum will give an
improved determination of $f(k_+)$, and
\item studying the dependence of the extracted value of $|V_{ub}|$ as a function
of the lepton cut $E_\ell$ can test the size of the subleading twist terms in
(\ref{Vubrelation}).
\end{itemize}

\begin{table*}[hbt]
\caption{Comparison of different kinematic cuts for the determination of
$|V_{ub}|$ from inclusive decays.}

\begin{tabular}{|c|c|l|l|}
\hline 
~Cut~ &  \% of rate  &  ~Good~  &  ~Bad~  \\ \hline
&&& - depends on $f(k^+)$ (and subleading\\
$E_\ell>{m_B^2-m_D^2\over 2 m_B}$ & $\sim 10\%$  & - simplest to measure &\ \
corrections) \\
&&&  - WA corrections may be substantial \\
&&&  - reduced phase space - duality issues? \\ \hline
$m_X<m_D$ & $\sim 70\%$  &  - lots of rate  &  - depends on $f(k^+)$ (and
subleading \\ 
&&&\ \ corrections)\\ \hline 
&&&- very sensitive to $m_b$\\ 
$q^2>(m_B-m_D)^2$ & $\sim 20\%$  & - insensitive to $f(k^+)$  &  - WA
corrections may be substantial \\
&&&- effective expansion parameter is \\
&&& \ \ $\Lambda_{\rm QCD}/m_c$ \\ \hline
&&- insensitive to $f(k^+)$  & \\
``Optimized cuts" & up to &- lots of rate  &   - less rate than pure $m_X$
cut,\\
&$\sim 45\%$ &- can move cuts away from  &   \ \ and more complicated to measure
   \\
&&\ \ kinematic limits and still have  &  \\
&&\ \ small uncertainties  &   \\
\hline
\end{tabular}\\[4pt]
\label{comparetable}
\end{table*}

\section{Spectral Moments}

Since differential rates may be computed in the HQE as a power series in
$\alpha_s(m_b)$ and $\Lambda_{\rm QCD}/m_b$, an unlimited number of spectral
moments may be computed.  Different moments have different dependence on the
nonperturbative parameters of the HQE, so a simultaneous fit to multiple moments
allows these parameters to be determined experimentally.  In addition,
consistency between different observables provides a powerful check of the
validity of the HQE to inclusive decays.   Moments which have received
particular attention are moments of the charged lepton energy spectrum
\cite{chargedleptons} and hadronic invariant mass spectrum
\cite{hadronicinvariantmass} in $B\to X_c\ell\bar\nu$ decays, and moments of the
photon energy spectrum in $B\to X_s\gamma$ decays \cite{btosgammaspectrum}.

By comparing the first moment of the photon spectrum in $B\to X_s\gamma$ with
the first moment of the hadronic invariant mass spectrum, the CLEO collaboration
\cite{cleomoments1} determined $\bar\Lambda=0.35\pm 0.07\pm 0.10\,\mbox{GeV}$
and $\lambda_1=-0.236\pm 0.071\pm 0.078\ \mbox{GeV}^2$ (where the first error is
experimental and the second theoretical).  More recently, additional moments
have been measured by CLEO\cite{CLEOnew},  BABAR \cite{Babarhadron} and DELPHI
\cite{DELPHImoments}, providing enough constraints to perform a global fit to
the HQE including terms of order $1/m_b^3$.  Two such global fits have been
recently performed.  Ref.\ \cite{bllm02} found
\begin{eqnarray}
m_b^{1S}&=&4.75\pm 0.10\,\rm{GeV}\\
V_{cb}&=&(40.8\pm 0.9)\times 10^{-3}\nonumber
\end{eqnarray}
while Ref.\ \cite{battaglia03} found, from just the DELPHI moments,
\begin{eqnarray}
m_b(1\ \rm{GeV})&=&4.59\pm 0.08\pm 0.01\,\rm{GeV} \\
m_c(1\ \rm{GeV})&=&1.13\pm 0.13\pm 0.03\,\rm{GeV}\nonumber\\ 
V_{cb}&=&(41.1\pm 1.1)\times 10^{-3}\nonumber
\end{eqnarray}
(corresponding to $m_b^{1S}=4.69$ GeV.)  Ref.\ \cite{battaglia03}  also
performed a fit using the pole mass scheme.  Matrix elements arising up to
$O(1/m_b^3)$ in the HQE were also determined from each of the fits.

In the approach of Ref.\ \cite{bllm02}, the charm quark mass is determined by
the heavy quark relation
\begin{eqnarray}
m_b - m_c &=& \bar m_B-\bar m_D  - \lambda_1 \left( \frac{1}{2m_c} -
\frac{1}{2m_b} \right)\\
  &&+ (\rho_1 - \tau_1 - \tau_3) \left( \frac{1}{4m_c^2} - \frac{1}{4m_b^2}
\right)+O\left({1\over m_{c,b}^3}\right)\nonumber
\end{eqnarray}
(where $\bar m_{B(D)}\equiv (m_{B(D)}+3 m_{B^*(D^*)})/4$ is the spin-averaged
meson mass, and $\rho_1$ and the $\tau_i$'s are matrix elements of order
$\Lambda_{\rm QCD}^3$).
In the approach of Ref.\ \cite{battaglia03},  the charm quark is not treated as
heavy, and its mass is taken to be a free parameter, to be fit from the moments.
 The second approach has the advantage that it does not require an expansion in
$\Lambda_{\rm QCD}/m_c$.  It however has the disadvantage that if the $c$ quark
is not treated as heavy, a systematic determination of the parameters
$\lambda_2$ and $\rho_2$ (arising at $O(1/m^2)$ and $O(1/m^3)$, respectively)
from the $D-D^*$ and $B-B^*$ mass splittings can no longer be performed.  

It is amusing to note that setting experimental errors in the global fit of
Ref.\ \cite{bllm02} to zero, one obtains the uncertainties $\delta(|V_{cb}|=\pm
0.35\times 10^{-3}$, $\delta(m_b)=\pm 35$ MeV, which give an indication of the
limiting theoretical uncertainty in the analysis.

One can also use this fit to make precise predictions of other moments as a
crosscheck.  Bauer and Trott \cite{BT02} examined fractional moments of the
lepton spectrum with a variety of cuts, and found certain moments that were very
insensitive to nonperturbative effects.   These moments may therefore be
predicted with small uncertainties using the values of $\bar\Lambda$ and
$\lambda_1$ determined from other moments, providing a stringent test of the HQE
for inclusive decays.  Later, these moments were measured by CLEO\cite{CLEOnew},
and were found to agree beautifully with the theoretical predictions:
\begin{eqnarray}
D_3&\equiv& {\int_{1.6\,{\rm GeV}} E_\ell^{0.7}{d\Gamma\over
dE_\ell}\,dE_\ell\over \int_{1.5\,{\rm GeV}} E_\ell^{1.5}{d\Gamma\over
dE_\ell}\,dE_\ell}=\cases{0.5190\pm 0.0007&(T)\cr
0.5193\pm 0.0008&(E)} \nonumber\\
D_4&\equiv& {\int_{1.6\,{\rm GeV}} E_\ell^{2.3}{d\Gamma\over
dE_\ell}\,dE_\ell\over \int_{1.5\,{\rm GeV}} E_\ell^{2.9}{d\Gamma\over
dE_\ell}\,dE_\ell}=\cases{0.6034\pm 0.0008&(T)\cr
0.6036\pm 0.0006&(E)} \nonumber\\
\end{eqnarray}
(where ``T" and ``E" denote theory and experiment, respectively).

\begin{figure}[ht]
\includegraphics[width=7cm]{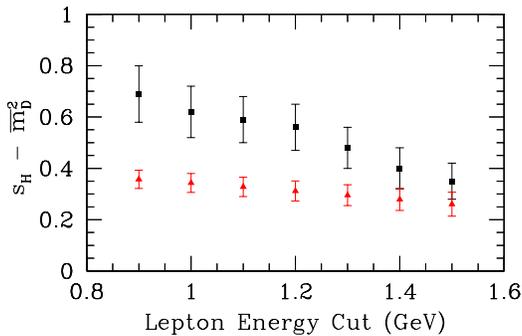}
\caption{Comparison of the BABAR measurements of the hadron invariant mass
spectrum vs. the lepton energy cut (black squares), and the HQE prediction not
including BABAR hadronic mass data (red triangles), from Ref.\ \cite{bllm02}.
\label{babarsH}}
\end{figure}

It should be noted, however, that there is currently poor agreement between the
HQE and the first hadronic invariant mass moment, measured by BABAR with
different lepton energy cuts, as shown in Fig.\  \ref{babarsH}.  This is a
worrisome feature which should be better understood.  One possible resolution is
to note that the BABAR measurement is not completely model independent, but
depends on the assumed spectrum of excited $D$ resonances.  In the model used,
there is no contribution to the $B$ semileptonic width from excited $D$ states
with masses below $\sim 2.4$ GeV.   This is an assumption which is in conflict
with the results of the HQE, as noted a number of years ago by Gremm and
Kapustin \cite{gk}. 

Denoting the lowest excited state as $D^{**}$, a lower bound on the first moment
of the hadronic invariant mass spectrum may be obtained
\cite{hadronicinvariantmass} by assuming that all rate which does not go to the
$D$ or $D^*$ goes to the $D^{**}$.  This gives the inequality
\begin{eqnarray}
\langle s_H-\bar m_D^2\rangle&=&\bar
m_B^2\left(0.051{\alpha_s\over\pi}+0.23{\bar\Lambda\over\bar m_B}+\dots\right)\\
&&\hskip-0.8in\geq \Gamma_{D^{**}}(m_{D^{**}}^2-\bar
m_D^2)+\Gamma_{D^*}(m_{D^*}^2
-\bar m_D^2)+\Gamma_D(m_D^2-\bar m_D^2)\nonumber
\end{eqnarray}
where $\Gamma_X\equiv \Gamma(B\to X\ell\bar\nu)/\Gamma_{S.L.}$ is the fraction
of the semileptonic width which goes to the final state $X$, and the first line
gives the first two terms in the HQE.  Combining the HQE prediction with the
measured ratio of the semileptonic $B\to D$ and $B\to D^*$ widths \cite{pdg}
\begin{equation}
\Gamma_D=0.31(1-\Gamma_{D^{**}}),\ \ \Gamma_{D^*}=0.69(1-\Gamma_{D^{**}})
\end{equation}
and taking $m_{D^{**}}=2.45$ GeV leads to the upper bound
\begin{equation}
\Gamma_{D^{**}}<0.22
\end{equation}
which is in conflict with the experimental semileptonic branching fraction to
excited states of $\sim0.35$.   Thus, if the prediction for the first moment of
the hadronic invariant mass spectrum is valid,  a non-negligible fraction of the
semileptonic $B$ width must be to excited $D$ states with $m_X<2.45$ GeV, which
would also bring the BABAR results in better agreement with theory
\cite{luthichep}.   It will be interesting to see how this situation evolves. 

\section{Conclusions}

The heavy quark expansion (HQE) has proven very successful at giving a
model-independent description of inclusive $B$ decays.
Theory and experiment are now at the stage that a precision 
determination of $|V_{ub}|$ is possible from inclusive decays.  The challenge is
to incorporate kinematic cuts that exclude $b\to c$ decays without introducing
large uncertainties (theoretical or experimental).  Cutting on the lepton
invariant mass $q^2$ or an optimized combination of $q^2$ and the hadronic
invariant mass $m_X$ gives a result that is insensitive to the nonperturbative
light-cone distribution function $f(k_+)$, at the expense of a difficult
experimental measurement.  Cutting on $m_X$ or the energy of the charged lepton
$E_\ell$ is easier experimentally, but introduces dependence on the
nonperturbative parton distribution function $f(k_+)$.  At leading order in
$1/m_B$, $f(k_+)$ may be determined from $B\to X_s\gamma$ decays, but there are
potentially large subleading corrections to this relation (at least for the cut
on $E_\ell$) which may limit the ultimate precision of this method.

Spectral moments from semileptonic $b\to c$ and radiative $b\to s$ decays are
now being used to study the HQE at the $O(1/m_b^3)$ level, and to determine the
values of the hadronic matrix elements which arise in the expansion.  Global
fits to a variety of moments now allow $V_{cb}$ to be determined with an
uncertainty at the $2\%$ level, and a short-distance $m_b$ with an uncertainty
at the 100 MeV level.  These determinations are not yet limited by theory, and
so these determinations are likely to improve in the future.

\end{document}